\documentclass[11pt]{article}

\usepackage{amsmath}
\usepackage{graphicx}
\usepackage{amsfonts}
\usepackage{amssymb}
\usepackage{epsfig}
\usepackage{color}
\usepackage{psfrag}
\usepackage{epstopdf}

\usepackage{caption}
\usepackage{subcaption}

\newcommand{\EQ}{\begin{equation}}
\newcommand{\EN}{\end{equation}}
\newcommand{\be}{\begin{equation}}
\newcommand{\ee}{\end{equation}}
\newcommand{\bea}{\begin{eqnarray}}
\newcommand{\eea}{\end{eqnarray}}

\begin{document} \setcounter{page}{0}
\topmargin 0pt
\oddsidemargin 5mm
\renewcommand{\thefootnote}{\arabic{footnote}}
\newpage
\setcounter{page}{0}
\topmargin 0pt
\oddsidemargin 5mm
\renewcommand{\thefootnote}{\arabic{footnote}}
\newpage
\begin{titlepage}
\begin{flushright}
\end{flushright}
\vspace{0.5cm}
\begin{center}
{\large {\bf Space of initial conditions and universality}\\
{\bf  in nonequilibrium quantum dynamics}}\\
\vspace{1.8cm}
{\large Gesualdo Delfino$^{1,2}$ and Marianna Sorba$^{1,2}$}\\
\vspace{0.5cm}
{\em $^1$SISSA -- Via Bonomea 265, 34136 Trieste, Italy}\\
{\em $^2$INFN sezione di Trieste, 34100 Trieste, Italy}\\
\end{center}
\vspace{1.2cm}

\renewcommand{\thefootnote}{\arabic{footnote}}
\setcounter{footnote}{0}

\begin{abstract}
\noindent
We study analytically the role of initial conditions in nonequilibrium quantum dynamics considering the one-dimensional ferromagnets in the regime of spontaneously broken symmetry. We analyze the expectation value of local operators for the infinite-dimensional space of initial conditions of domain wall type, generally intended as initial conditions spatially interpolating between two different ground states. At large times the unitary time evolution takes place inside a light cone produced by the spatial inhomogeneity of the initial condition. In the innermost part of the light cone the form of the space-time dependence is universal, in the sense that it is specified by data of the equilibrium universality class. The global limit shape in the variable $x/t$ changes with the initial condition. In systems with more than two ground states the tuning of an interaction parameter can induce a transition which is the nonequilibrium quantum analog of the interfacial wetting transition occurring in classical systems at equilibrium. We illustrate the general results through the examples of the Ising, Potts and Ashkin-Teller chains. 
\end{abstract}
\end{titlepage}

\newpage
\tableofcontents
\section{Introduction}
Universality is a central paradigm of statistical physics. In the equilibrium setting, in which it originated and is well established \cite{WK}, it says that systems possessing a continuous phase transition exhibit a number of quantitative properties that do not depend on the microscopic details of the Hamiltonian, for example the range of the interaction as long as it is short. In a given space dimensionality, these {\it universal} properties depend instead on the group $G$ of internal transformations that leave the Hamiltonian invariant, so that  $G$ is the label of the different universality classes. 

The extension of such a notion of universality to the nonequilibrium framework is a nontrivial task. In the quantum case that we consider in this paper, a time independent Hamiltonian $H$ generates unitary time evolution as in equilibrium. However, observables are now expectation values on a "nonequilibrium state" $|\psi\rangle$ that is not an eigenstate of $H$, but rather, in the physically interesting cases, a superposition of infinitely many eigenstates. The question then arises of whether the notion of universality is compatible with the presence of the infinitely many coefficients of the superposition, which in turn correspond to infinitely many possible initial conditions of the time evolution. Since at the mathematical level these coefficients can be arbitrary, it is essential to refer to well defined physical problems. 

In one of these, the Hamiltonian $H$ that rules the time evolution for time $t>0$ results from the change of an interaction parameter at $t=0$ \cite{BMcD}. The name "quantum quench" \cite{SPS,CC} has been introduced for this problem in analogy with thermal quenches in classical systems. If for $t<0$ the system was in the ground state of the pre-quench Hamiltonian $H_0$, the nonequilibrium post-quench state $|\psi\rangle$ is dynamically generated, with the coefficients of the superposition entirely determined by the quench. The analytical determination of $|\psi\rangle$ is nontrivial, but the general formalism has been developed in the last years \cite{quench,DV_quench,oscill,oscillD}. It shows, in particular, how $|\psi\rangle$ depends on the equilibrium universality class and how the transformation properties of the quench operator under the group $G$ affect the dynamics at large times. 

A physical problem involving no quench but still leading to a nonequilibrium state $|\psi\rangle$ is that in which $H$ is the Hamiltonian of a translation invariant system, but the initial condition of the time evolution is spatially inhomogeneous. In this paper we will study this problem for the case of the one-dimensional ferromagnets with interaction parameters in the range in which the spontaneous breaking of a discrete symmetry $G$ leads to degenerate ground states that we denote $|0_a\rangle$, $a=1,2,\ldots,N$. We are interested in the expectation value $\langle\Phi(x,t)\rangle_{ab}$ of a local operator $\Phi$ (e.g. the order parameter operator), for initial ($t=0$) conditions that interpolate between a ground state $|0_a\rangle$ as $x\to -\infty$, and a different ground state $|0_b\rangle$ as $x\to +\infty$. This interpolation is chosen to preserve the symmetry $G$ of the system, but still can be realized in infinitely many ways, meaning that the corresponding nonequilibrium states $|\psi\rangle$ form an infinite-dimensional space ${\cal W}$. We refer to ${\cal W}$ as the space of domain wall states or, equivalently, of domain wall initial conditions. We will show that for any initial condition belonging to a subspace ${\cal W}_1$ -- itself infinite dimensional -- of ${\cal W}$ the time evolution leads for large $t$ to\footnote{Throughout the paper the symbol $\simeq$ indicates omission of terms subleading for large $t$.} 
\EQ
\langle\Phi(x,t)\rangle_{ab}\simeq\left\{
\begin{array}{l}
\langle\Phi\rangle_a\,,\hspace{.5cm}x<-t\,,\\
\\
\frac{1}{2}\left[\langle\Phi\rangle_a+\langle\Phi\rangle_b\right]+{\cal A}
\left\{C_0^\Phi/M-\left[\langle\Phi\rangle_a-\langle\Phi\rangle_b\right]\,x\right\}t^{-1}\,,\hspace{.6cm}{|x|}\ll t\,,\\
\\
\langle\Phi\rangle_b\,,\hspace{.5cm}x>t\,,
\end{array}
\right.
\label{results}
\EN
where
\EQ
\langle\Phi\rangle_a=\langle 0_a|\Phi(x,t)|0_a\rangle
\label{vev}
\EN
is the equilibrium expectation value  in the ground state $|0_a\rangle$, $C_0^\Phi$ is also determined by the equilibrium universality class, $M$ is the equilibrium mass gap, and the positive dimensionless amplitude ${\cal A}$ is the only quantity depending on the specific initial condition. The flat results for $|x|>t$ show the presence of a light cone expanding with velocity 1, which in our natural units is the maximal velocity of the excitation modes of the system\footnote{The presence of a light cone was originally observed in the study of a free fermionic chain with a steplike initial condition \cite{ARRS}.}. Concerning the region inside the light cone ($|x|<t$), our derivation will make explicit that the generality of the result for $|x|\ll t$ comes from the fact that in this limit the excitation modes with largest wavelength dominate, and these are maximally insensitive to the details of the initial condition. On the other hand, we also show that if the operator distinguishes the two ground states, namely if $\langle\Phi\rangle_a\neq\langle\Phi\rangle_b$, when $t\to\infty$ $\langle\Phi(x,t)\rangle_{ab}$ tends everywhere to a function of $x/t$ changing with the initial condition. Since the space ${\cal W}_1$ includes the low energy modes relevant for the region $|x|\ll t$, we expect the result (\ref{results}) to generally hold at large times, and explicitly illustrate how this extension occurs. A caveat arises in systems with at least three degenerate ground states when the change of an interaction parameter causes an "unbinding" transition to a regime in which ${\cal W}_1$ shrinks to zero and the result (\ref{results}) is affected for $|x|\ll t$. This is the same mechanism \cite{DV,DS} responsible for interfacial wetting \cite{Dietrich} in the theory of phase separation in classical systems at equilibrium. 

It is worth pointing out that the problem of identifying and analytically determining universal properties emerging at late times in nonequilibrium quantum dynamics -- irrespectively of the fine details of the initial condition -- is addressed here for the first time. The theoretical advances we achieve are made possible by our ability to cast the problem in the nonperturbative field theoretical framework, the natural one in which to address the questions concerning universality. For the problem at hand, it allows us to generally perform the analysis for all the universality classes displaying spontaneous symmetry breaking in one spatial dimension, and for the infinite-dimensional space of initial conditions corresponding to the same topological class of nonequilibrium states spatially interpolating between different ferromagnetic ground states.
 
The paper is organized as follows. The main analysis is performed in the next section, with a generalization presented in the appendix. The results are then illustrated in section~3 through the examples of the Ising, Potts and Ashkin-Teller chains, while section~4 contains few concluding remarks.

\section{Space of domain wall initial conditions and time evolution}
\label{sec_main}
The elementary excitation modes of a one-dimensional ferromagnet in the regime of spontaneously broken symmetry are kinks interpolating between two degenerate ground states. In the proximity of the quantum critical point these are relativistic quasiparticles with energy and momentum
\EQ
(E,p)=(M\cosh\theta,M\sinh\theta)\,,
\EN
where $\theta$ is called rapidity, and the mass $M$ is a measure of the deviation from criticality. The interpolating initial conditions we are interested in correspond to nonequilibrium states of the form
\EQ
|\psi\rangle=\sum_{n=1}^\infty |\psi_n\rangle=\sum_{n=1}^\infty \int d\theta_1\ldots d\theta_n\,f_n(\theta_1,\ldots,\theta_n)\,|\theta_1,\ldots,\theta_n\rangle\,,
\label{psi}
\EN
where $|\theta_1,\ldots,\theta_n\rangle$ is a $n$-kink state starting in $|0_a\rangle$ and ending in\footnote{It is understood that for $N>2$ degenerate ground states and $n>1$ the expansion (\ref{psi}) includes a sum over the intermediate ground states visited in the $n$-step path from $|0_a\rangle$ to $|0_b\rangle$.} $|0_b\rangle$. The different choices of the functions $f_n$ span the space ${\cal W}$ of the domain wall states $|\psi\rangle$ and allow for arbitrary spatial interpolation in the initial condition. Since we consider initial conditions that do not introduce any explicit breaking of the symmetry of the system under the group $G$, the functions $f_n$ are required to preserve this property. They are also required to decay for $\theta_i\to\pm\infty$ sufficiently rapidly to ensure convergence of the integrals over rapidities, and to be free of singularities for real values of the rapidities. The expectation value of a local operator with such a general symmetry preserving domain wall initial condition is given by
\EQ
\langle\Phi(x,t)\rangle_{ab} = \frac{\langle\psi|\Phi(x,t)|\psi\rangle}{\langle\psi|\psi\rangle}\,.
\label{full}
\EN

We will now investigate the properties of the dynamics in the infinite dimensional subspace ${\cal W}_1$ of initial conditions corresponding to one-kink states
\EQ
|\psi_1\rangle=\int d\theta\,f(\theta)\,|\theta\rangle\,,
\label{psi1}
\EN
where we have simplified the notation setting $f_1=f$. In this subspace the expectation value (\ref{full}) reads
\bea
G_\Phi(x,t) &=& \frac{1}{N_f}\,\langle\psi_1|\Phi(x,t)|\psi_1\rangle
\label{Phi_ab}\\
&=& \frac{1}{N_f}\int d\theta_1 d\theta_2\,f^*(\theta_1)f(\theta_2)\,F_\Phi(\theta_1-\theta_2)\,
e^{i[(p_1-p_2)x+(E_1-E_2)t]}\,,\nonumber
\eea
where we defined\footnote{We use the state normalization $\langle\theta|\theta'\rangle=2\pi\,\delta(\theta-\theta')$.}
\EQ
N_f=\langle\psi_1|\psi_1\rangle=2\pi\int d\theta\,|f(\theta)|^2\,,
\label{Nf}
\EN
\EQ
F_\Phi(\theta_1-\theta_2)=\langle\theta_1|\Phi(0,0)|\theta_2\rangle\,,
\label{F}
\EN
and used
\EQ
\Phi(x,t)=e^{i(Px+Ht)}\,\Phi(0,0)\,e^{-i(Px+Ht)}\,,
\EN
with $P$ the momentum operator. The operators $\Phi$ of our interest are invariant under relativistic transformations and, since such a transformation shifts rapidities by a constant, the matrix element (\ref{F}) depends on the rapidity difference. It can be written as \cite{DC98}
\EQ
F_\Phi(\theta)=i\frac{\langle\Phi\rangle_a-\langle\Phi\rangle_b}{\theta-i\epsilon}+\sum_{k=0}^\infty C_k^\Phi\,\theta^k+2\pi\,\delta(\theta)\langle\Phi\rangle_a\,,
\label{expansion}
\EN
where the term containing $\delta(\theta)$ is the disconnected part corresponding to the annihilation of the particle on the left with the particle on the right, while the connected part has been expanded in powers of $\theta$. The pole term is a remnant in the connected part of the annihilation configuration\footnote{The pole is known to account for phase separation in the classical case at equilibrium \cite{DV}, in which it yields, in particular, the exact order parameter profile originally obtained in \cite{AR} from the lattice solution of the two-dimensional Ising model. Annihilation poles are well known in the multiparticle matrix elements of integrable theories \cite{Smirnov}, but integrability plays no role in the determination of the residue for the matrix element (\ref{F}) \cite{DC98}.} $\theta=0$, and the infinitesimal imaginary part $i\epsilon$ specifies the regularization prescription for the integral in (\ref{Phi_ab}). 

Let us call $G_\Phi^\textrm{sing}$ the contribution to (\ref{Phi_ab}) of (\ref{expansion}) without the regular part $\sum_{k=0}^\infty C_k^\Phi\,\theta^k$. Defining $\theta_\pm=\theta_1\pm\theta_2$ we write this contribution in the form
\EQ
G_\Phi^\textrm{sing}(x,t)=\langle\Phi\rangle_a+i\frac{\langle\Phi\rangle_a-\langle\Phi\rangle_b}{2N_f}\int d\theta_+ d\theta_-\, \frac{f^*(\theta_1)f(\theta_2)}{\theta_- -i\epsilon}\,e^{2iMt\,B(x/t,\theta_+)\sinh\frac{\theta_-}{2}},
\label{Gsing}
\EN
where
\EQ
B(x/t,\theta_+)=\frac{x}{t}\cosh\frac{\theta_+}{2}+\sinh\frac{\theta_+}{2}\,.
\label{B}
\EN
We can set $\sinh\frac{\theta_-}{2}=p$ and consider the integral over $p$ in which we close the contour in the upper (lower) complex half-plane if $B$ is positive (negative). In particular, we can close the contour along the line with constant imaginary part $\textrm{Im}\,p=c$. When $t\to\infty$, the contribution coming from the integral on this line is suppressed as $e^{-2M|cB|t}$ and can be neglected. On the other hand, we can reduce $|c|$ in such a way that the closed integration contour contains only the singularity at $p=i\epsilon/2$ for $c>0$, and no singularity at all for $c<0$. Hence, Cauchy's residue integration tells us that for $t$ large 
\EQ
G_\Phi^\textrm{sing}(x,t)\simeq\langle\Phi\rangle_a-\frac{\langle\Phi\rangle_a-\langle\Phi\rangle_b}{N_f}\,2\pi\int_{\theta_0}^\infty d\theta\,|f(\theta)|^2\,,
\label{singular}
\EN
where $\theta_0$ is the value of $\theta$ above which $\tanh\theta>-x/t$. Since $\theta_0$ is equal to $+\infty$ when $x/t<-1$ and to $-\infty$ when $x/t>1$, we have
\EQ
G_\Phi^\textrm{sing}(x,t)\simeq\left\{
\begin{array}{l}
\langle\Phi\rangle_a\,,\hspace{.5cm}x<-t\,,\\
\\
\langle\Phi\rangle_b\,,\hspace{.5cm}x>t\,
\end{array}
\right.
\EN
for $t$ large. On the other hand, $\theta_0\simeq -x/t$ when $|x|/t\ll 1$, and we can break the integration over $\theta$ into that on the small interval between $-x/t$ and $0$, in which $f(\theta)\simeq f(0)$, and that for $\theta>0$. Since the ground states $|0_a\rangle$ and $|0_b\rangle$ are exchanged by the symmetry $G$ of the Hamiltonian and play a symmetric role preserved by the initial condition, $|f(\theta)|^2$ is an even function. As a consequence $2\pi\int_0^\infty d\theta\,|f(\theta)|^2=N_f/2$ and we have
\EQ
G_\Phi^\textrm{sing}(x,t)\simeq\frac{\langle\Phi\rangle_a+\langle\Phi\rangle_b}{2}-2\pi A_f
\left[\langle\Phi\rangle_a-\langle\Phi\rangle_b\right]\,\frac{x}{t}\,,\hspace{.9cm}|x|\ll t\,,
\EN
with $A_f=|f(0)|^2/N_f$.

Now we consider the contribution to (\ref{Phi_ab}) coming from the regular part $\sum_{k=0}^\infty C_k^\Phi\,\theta^k$ of (\ref{expansion}), namely
\EQ
G_\Phi^\textrm{reg}(x,t)=\frac{1}{N_f}\sum_{k=0}^\infty C_k^\Phi \int d\theta_1 d\theta_2\,f^*(\theta_1)f(\theta_2)\,(\theta_1-\theta_2)^k\,e^{i[(p_1-p_2)x+(E_1-E_2)t]}\,.
\label{regular}
\EN
We first observe that for $t$ large enough the rapid oscillations of the integrand suppress the integral unless the phase is stationary, namely unless $E_j x+p_j t=0$, $j=1,2$. Since $|p_j|/E_j=|\tanh\theta_j|<1$, the stationarity condition is satisfied inside the light cone\footnote{This mechanism leading to the light cone can be compared with that for two-point functions in homogeneous systems out of equilibrium, in which the connectedness structure of matrix elements plays an essential role \cite{2point_spreading}.} $|x|<t$. Hence, for $t$ large we have
\EQ
G_\Phi^\textrm{reg}(x,t)\simeq 0\,,\hspace{1cm}|x|>t\,,
\label{reg_cone}
\EN
where the corrections are small and rapidly vanishing as $|x|$ increases with $t$ fixed. 

On the other hand, the stationarity condition $\tanh\theta_j=-x/t$ implies that inside the light cone small rapidities dominate the integral for $|x|/t\ll 1$. Hence in this limit we can write
\EQ
G_\Phi^\textrm{reg}(x,t)\simeq A_f\sum_{k=0}^\infty C_k^\Phi \int d\theta_1 d\theta_2\,(\theta_1-\theta_2)^k\,e^{iM[(\theta_1-\theta_2)x+\frac{1}{2}\theta_1^2(t+i\epsilon)-\frac{1}{2}\theta_2^2(t-i\epsilon)]}\,,
\label{Greg}
\EN
where the infinitesimal imaginary parts added to $t$ preserve the convergence of the integral. We can now rescale the rapidities and deduce that the $k$-th term decays at large times as $t^{-(k+2)/2}$. Hence the leading contribution comes from $k=0$ and we have
\bea
G_\Phi^\textrm{reg}(x,t) &\simeq & A_f\,C_0^\Phi \left|\int d\theta\,e^{iM[\theta x+\frac{1}{2}\theta^2 (t+i\epsilon)]}\right|^2=A_f\,C_0^\Phi \frac{2\pi}{M|t+i\epsilon|}\,e^{-\frac{M\epsilon x^2}{t^2+\epsilon^2}}\nonumber\\
&\to & A_f\,C_0^\Phi\,\frac{2\pi }{Mt}\,,
\hspace{1cm}|x|\ll t\,.
\label{regular_ir}
\eea

Putting all together, since $G_\Phi(x,t)$ is the sum of $G_\Phi^\textrm{sing}(x,t)$ and $G_\Phi^\textrm{reg}(x,t)$, the results that we obtained for these two terms lead to the large time behaviors (\ref{results}), where 
\EQ
{\cal A}=2\pi A_f=2\pi\,\frac{|f(0)|^2}{N_f}\,.
\label{A}
\EN

When $\langle\Phi\rangle_a\neq\langle\Phi\rangle_b$, $G_\Phi(x,t)$ should tend as $t\to\infty$ to a nonconstant limit shape as a function of $x/t$, since in this variable the edges of the light cone are fixed at $\pm 1$. To see this, we notice that if the large time analysis of the contribution (\ref{regular}) of regular terms is performed at a generic point $x$ inside the light cone, the stationarity condition $\tanh\theta_j=-x/t$ selects the rapidities around which to expand to evaluate the integral, and considerations analogous to those we just made for $x$ small lead to the conclusion that again this contribution is suppressed at large times. It follows that (\ref{regular}) goes to zero for any $x$ at large times, so that in this limit the dominant $x$-dependence is given by (\ref{singular}), as long as $\langle\Phi\rangle_a\neq\langle\Phi\rangle_b$. Hence, for $t\to\infty$ and $\langle\Phi\rangle_a\neq\langle\Phi\rangle_b$, $G_\Phi(x,t)\to G_\Phi^\textrm{sing}(x,t)$ and is a function of $x/t$, since $\theta_0$ in (\ref{singular}) depends on $x/t$. We also see that the large time limit of $G_\Phi(x,t)/\langle\Phi\rangle_a$ changes with the initial condition and depends on the universality class only through the "dilatation" factor $\langle\Phi\rangle_b/\langle\Phi\rangle_a$.


We saw that the generality of the large time result (\ref{results}) for $|x|\ll t$ is due to the dominance in this region of the low energy modes\footnote{The fact that the low energy modes dominate the large time dynamics in the general case of interacting quasiparticles is known for quantum quenches \cite{quench,DV_quench,oscill,oscillD}.}. Since the energy of a $n$-kink state is at least $nM$, this result should generically\footnote{Fine tuning of the functions $f_n$ can lead to peculiar states. These, however, will form some zero measure subset, and typically will not be physically relevant.} hold in the whole space ${\cal W}$ of domain wall states, as long as the one-kink contribution from ${\cal W}_1$ is present, and we show in the appendix the mechanism through which this extension occurs. The one-kink state will naturally be present in the states (\ref{psi}) arising in physical applications, but an interesting exception occurs for systems with more than two degenerate ground states, where the tuning of an interaction parameter can induce a transition to a regime in which ${\cal W}_1$ is empty; we illustrate this phenomenon in section~\ref{Ashkin}.

\begin{figure}[t]
    \centering
    \begin{subfigure}[h]{0.45\textwidth}
        \includegraphics[width=\textwidth]{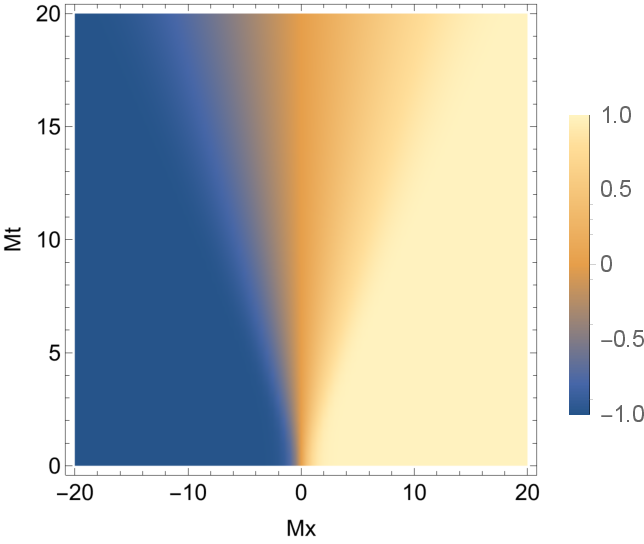}
    \end{subfigure}\hspace{1cm}%
    \begin{subfigure}[h]{0.45\textwidth}
        \includegraphics[width=\textwidth]{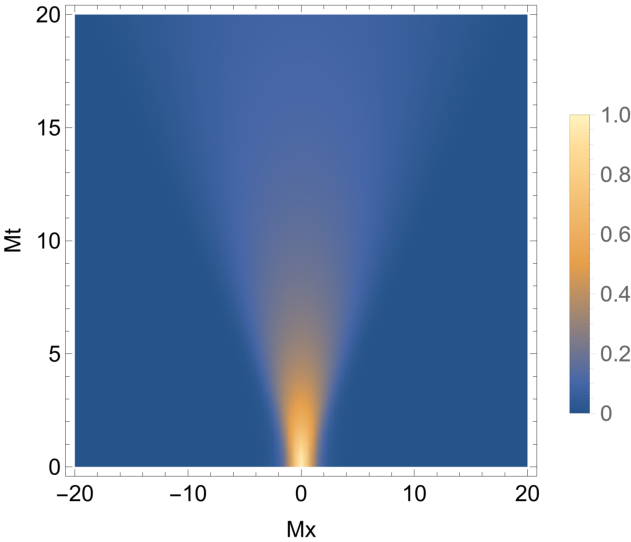}
    \end{subfigure}
    \caption{Ising magnetization components in the nonequilibrium state (\ref{gauss}) with $MR=1$.   
\textbf{Left:} Order parameter $\langle \sigma^x(x,t)\rangle_{-+}/\langle\sigma^x\rangle_+$. 
\textbf{Right:} Connected transverse magnetization $\langle\sigma^z(x,t)\rangle_{-+}^c/\langle\sigma^z(0,0)\rangle_{-+}^c$, with $\langle\sigma^z(x,t)\rangle_{-+}^c=\langle\sigma^z(x,t)\rangle_{-+}-\langle\sigma^z\rangle_{+}$.\\
    }
    \label{ising_density}
\end{figure}

\section{Some universality classes}
\subsection{Ising}
A first illustration of the results of the previous section is provided by the Ising universality class (symmetry group $G=\mathbb{Z}_2$), whose simplest lattice realization is the nearest neighbor transverse field Ising chain with Hamiltonian
\EQ
H_\textrm{Ising}=-J\sum_i\left(\sigma^x_i\sigma^x_{i+1}+g\,\sigma^z_i\right)\,,
\label{ising}
\EN
where $\sigma^{x,z}_i$ are Pauli matrices at site $i$, and $J>0$, $|g|<1$ is the ferromagnetically ordered regime of our interest. Denoting the two degenerate ground states as $|0_+\rangle$ and $|0_-\rangle$, we have $\langle\sigma^x\rangle_-=-\langle\sigma^x\rangle_+$ and $\langle\sigma^z\rangle_-=\langle\sigma^z\rangle_+$. Hence, for the order parameter operator $\sigma^x$, the large time result (\ref{results}) becomes
\EQ
\frac{\langle\sigma^x(x,t)\rangle_{-+}}{\langle\sigma^x\rangle_+}\simeq\left\{
\begin{array}{l}
-1\,,\hspace{.5cm}x<-t\,,\\
\\
2{\cal A}
\,x/t\,,\hspace{.6cm}{|x|}\ll t\,,\\
\\
1\,,\hspace{.5cm}x>t\,.
\end{array}
\right.
\label{ising_profile}
\EN
Here we also took into account that $C_0^{\sigma^x}=0$, as expected on symmetry grounds and explicitly following from \cite{BKW} (see \cite{review} for a review) 
\EQ
F_{\sigma^x}(\theta)=[i\,\textrm{coth}(\theta/2)+2\pi\,\delta(\theta)]\,\langle\sigma^x\rangle_-\,.
\EN
Recalling our result that ${\cal A}$ depends on the initial condition, (\ref{ising_profile}) is consistent with the behavior displayed by the plots of \cite{ZGEN,EME} for the chain (\ref{ising}) with two different realizations of sharp (steplike) domain wall initial conditions. Similarly, our result that (\ref{ising_profile}) tends at large times to a function of $x/t$ depending on the initial condition explains the observations of \cite{EME} about the plots against $x/t$. 

\begin{figure}[t]
    \centering
    \begin{subfigure}[h]{0.45\textwidth}
        \includegraphics[width=\textwidth]{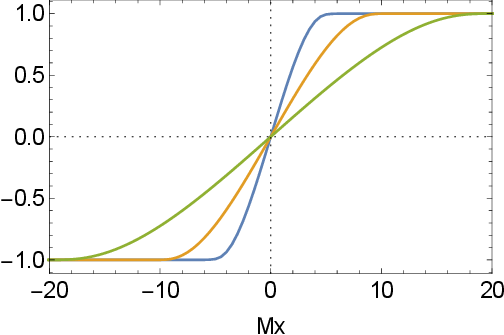}
    \end{subfigure}\hspace{1cm}%
    \begin{subfigure}[h]{0.45\textwidth}
        \includegraphics[width=\textwidth]{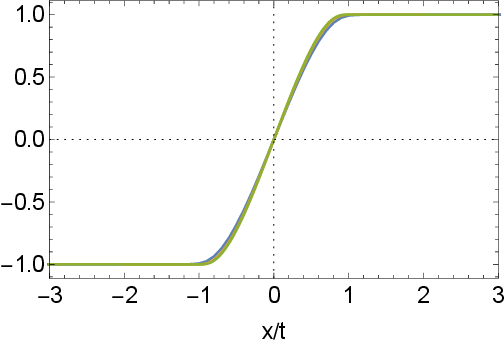}
    \end{subfigure}
    \caption{Ising order parameter $\langle \sigma^x(x,t)\rangle_{-+}/\langle\sigma^x\rangle_+$ in the nonequilibrium state (\ref{gauss}) with $MR=1$.  \textbf{Left:} $Mt=5,\,10,\,20$, in order of decreasing slope at $x=0$. 
\textbf{Right:} In the variable $x/t$ the three curves approach the large time limit given by (\ref{singular}). }
    \label{ising_op}
\end{figure}

We already remarked that for $|x|\ll t$ the dependence on the initial condition in (\ref{results}) is limited to the constant ${\cal A}$ because long wavelength  modes dominate in this region. On the other hand, when moving towards the edges of the light cone from inside, the fine structure of the initial condition becomes more and more relevant. For the chain (\ref{ising}) this feature is illustrated by the modulated behavior\footnote{This type of behavior was originally observed in a free fermion chain with sharp domain wall initial condition \cite{HRS}.} of the order parameter observed in \cite{ZGEN} zooming in close to the edges of the light cone for the two sharp domain wall initial conditions.

For the transverse magnetization, the result \cite{BKW,review}
\EQ
F_{\sigma^z}(\theta)=[{\cal C}\,\textrm{cosh}(\theta/2)+2\pi\,\delta(\theta)]\,\langle\sigma^z\rangle_+\,,
\EN
with ${\cal C}$ real and dimensionless, gives $C_0^{\sigma^z}={\cal C}\langle\sigma^z\rangle_+$, and then
\EQ
\frac{\langle\sigma^z(x,t)\rangle_{-+}}{\langle\sigma^z\rangle_+}\simeq\left\{
\begin{array}{l}
1\,,\hspace{.5cm}|x|>t\,,\\
\\
1+{\cal A}\,{\cal C}/(Mt)\,,\hspace{.6cm}{|x|}\ll t\,.\\
\end{array}
\right.
\EN 

\begin{figure}[t]
    \centering
    \begin{subfigure}[h]{0.45\textwidth}
        \includegraphics[width=\textwidth]{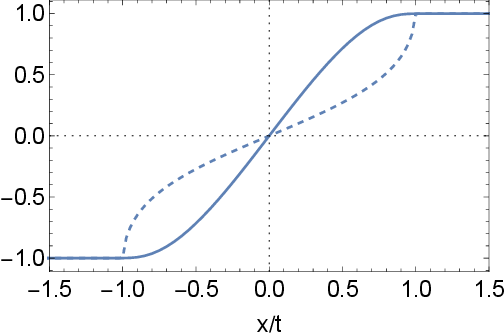}
    \end{subfigure}\hspace{1cm}%
    \begin{subfigure}[h]{0.45\textwidth}
      \includegraphics[width=\textwidth]{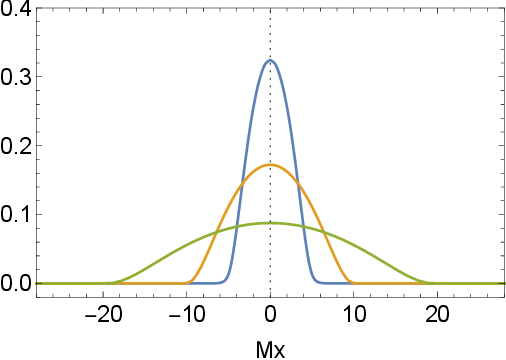}
    \end{subfigure}
    \caption{Ising magnetization components in the nonequilibrium state (\ref{gauss}). \textbf{Left:} Large time limit (\ref{singular}) of the order parameter $\langle \sigma^x(x,t)\rangle_{-+}/\langle\sigma^x\rangle_+$ for $MR=1$ (continuous line) and $MR=0.1$ (dashed line). \textbf{Right:} Connected transverse magnetization $\langle\sigma^z(x,t)\rangle_{-+}^c/\langle\sigma^z(0,0)\rangle_{-+}^c$ (as defined in fig.~\ref{ising_density}) for $MR=1$ and $Mt=5,\,10,\,20$, in order of decreasing value at $x=0$.}
    \label{ising_mixed}
\end{figure}

We show in figs.~\ref{ising_density}, \ref{ising_op} and \ref{ising_mixed} the expectation values $\langle\sigma^x(x,t)\rangle_{-+}$ and $\langle\sigma^z(x,t)\rangle_{-+}$ for the state (\ref{psi1}) with $f(\theta)=e^{-MR\,\theta^2}$, namely for the Gaussian wave packet
\EQ
\int d\theta\,e^{-MR\,\theta^2}\,|\theta\rangle\,.
\label{gauss}
\EN
The distance over which the order parameter significantly differs from the asymptotic values in the initial condition grows with $R$. The large time regime corresponds to $t$ much larger than $1/M$ and $R$. The state (\ref{gauss}) conveniently illustrates some global features of the large time behavior as the initial condition varies. In particular, the change in the order parameter limit shape shown in the left panel of fig.~\ref{ising_mixed} is essentially due to the fact that ${\cal A}$, and then the slope in the origin, decreases with $MR$. 

\subsection{Potts and the third phase}
\label{sec_potts}
The three-state Potts universality class, characterized by invariance under the permutational group $G=S_3$, finds its simplest representative in the nearest neighbor chain \cite{Solyom}
\EQ
H_\textrm{Potts}=-J\sum_i\left[\sigma_i^\dagger\sigma_{i+1}+\sigma_i\sigma_{i+1}^\dagger+g(M_i+M_i^\dagger)\right]\,,
\label{potts}
\EN
where $\sigma_i$ and $M_i$ are $3\times 3$ matrices satisfying $\sigma_i^2=\sigma_i^\dagger$, $\sigma_i^3=M_i^3=1$, $M_i^2=M_i^\dagger$, and $M_i\sigma_i=\omega\,\sigma_iM_i$, where $\omega=e^{2i\pi/3}$. Explicit representations are \
\EQ
\sigma=\left(\begin{array}{l}
1\,\,\,0\,\,\,0\\
0\,\,\,\omega\,\,\,0\\
0\,\,\,0\,\,\,\omega^2\\
\end{array}
\right)\,,
\hspace{1.5cm}
M=\left(\begin{array}{l}
0\,\,\,1\,\,\,0\\
0\,\,\,0\,\,\,1\\
1\,\,\,0\,\,\,0\\
\end{array}
\right)\,.
\EN
We refer to the ferromagnetically ordered regime $J>0$, $|g|<1$, in which there are three degenerate ground states $|0_a\rangle$, $a=1,2,3$. The Hermitian order parameter operator with components
\EQ
\sigma_a=\omega^{-a}\,\sigma+\omega^a\,\sigma^\dagger\,,\hspace{1cm}a=1,2,3\,,
\EN
satisfies $\sum_{a=1}^3\sigma_a=0$ and, by permutational symmetry,
\EQ
\langle\sigma_a\rangle_b=\frac{1}{2}\,(3\delta_{ab}-1)\,\langle\sigma_a\rangle_a\,.
\label{potts_vev}
\EN

\begin{figure}[t]
    \centering
    \begin{subfigure}[h]{0.45\textwidth}
        \includegraphics[width=\textwidth]{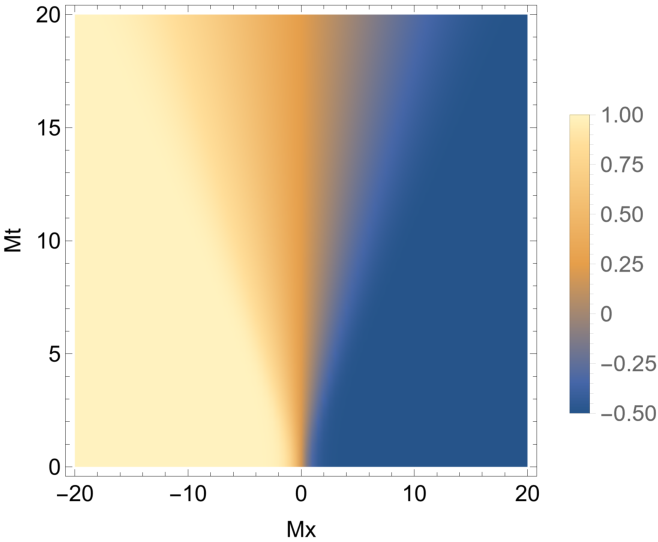}
    \end{subfigure}\hspace{1cm}%
    \begin{subfigure}[h]{0.45\textwidth}
        \includegraphics[width=\textwidth]{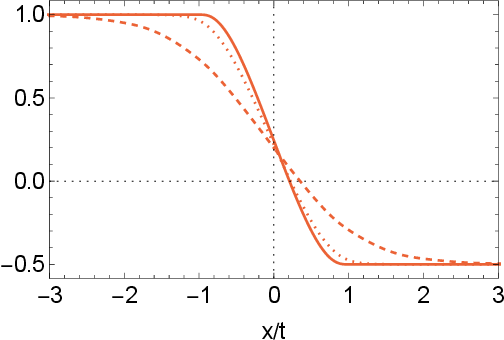}
    \end{subfigure}
    \caption{Potts order parameter component $\langle\sigma_1(x,t)\rangle_{12}/\langle\sigma_1\rangle_{1}$ in the nonequilibrium state (\ref{gauss}) with $MR=1$. On the right $Mt=1$ (dashed), $Mt=3$ (dotted) and $Mt=20$ (continuous); the latter curve is indistinguishable from the large time limit shape (\ref{singular}).}
\label{potts_1}
\end{figure}

\noindent
Hence, $\sigma_a$ detects phase $a$ and does not distinguish between the other two phases. The matrix elements \cite{DC98}
\EQ
F_{\sigma_1}(\theta)=\left[-\frac{\sqrt{3}}{2}\, \frac{\sinh\left(\frac{\theta}{6}-\frac{i\pi}{3}\right)}{\sinh\frac{\theta}{2}}\, {\cal F}(\theta)+2\pi \delta(\theta)\right]\langle\sigma_1\rangle_{1}\,,
\label{sigma1}
\EN
and
\EQ
F_{\sigma_3}(\theta)=\left[\frac{\sqrt{3}}{2}\,\frac{\sinh\left(\frac{\theta}{6}-\frac{i\pi}{3}\right)+\sinh\left(\frac{\theta}{6}+\frac{i\pi}{3}\right)}{\sinh\frac{\theta}{2}}\, {\cal F}(\theta)-\pi \delta(\theta)\right] \langle\sigma_1\rangle_{1}\,,
\label{sigma3}
\EN
where
\EQ
{\cal F}(\theta)= \exp\left\{\int_0^{\infty} dx\, \frac{2 \sinh\frac{2x}{3}}{x \sinh^2 x}\, \sin^2\frac{\theta x}{2\pi} \right\}\,,
\EN
determine through (\ref{expansion}) $C_0^{\sigma_1}=-\langle \sigma_1\rangle_1/(4\sqrt{3})$ and $C_0^{\sigma_3}=\langle \sigma_1\rangle_1/(2\sqrt{3})$, so that we have the large time behaviors
\EQ
\frac{\langle\sigma_1(x,t)\rangle_{12}}{\langle \sigma_1\rangle_1}\simeq\left\{
\begin{array}{l}
1\,,\hspace{.5cm}x<-t\,,\\
\\
1/4-{\cal A}\left(\frac{1}{4\sqrt{3}}+\frac{3}{2}Mx\right)/(Mt)\,,\hspace{.6cm}{|x|}\ll t\,,\\
\\
-1/2\,,\hspace{.5cm}x>t\,,\\
\end{array}
\right.
\EN
and
\EQ
\frac{\langle\sigma_3(x,t)\rangle_{12}}{\langle \sigma_3\rangle_1}\simeq\left\{
\begin{array}{l}
1\,,\hspace{1cm}|x|>t\,,\\
\\
1-{\cal A}/(\sqrt{3}Mt)\,,\hspace{.6cm}{|x|}\ll t\,.\\
\end{array}
\right.
\label{third}
\EN

\begin{figure}[t]
    \centering
    \begin{subfigure}[h]{0.45\textwidth}
        \includegraphics[width=\textwidth]{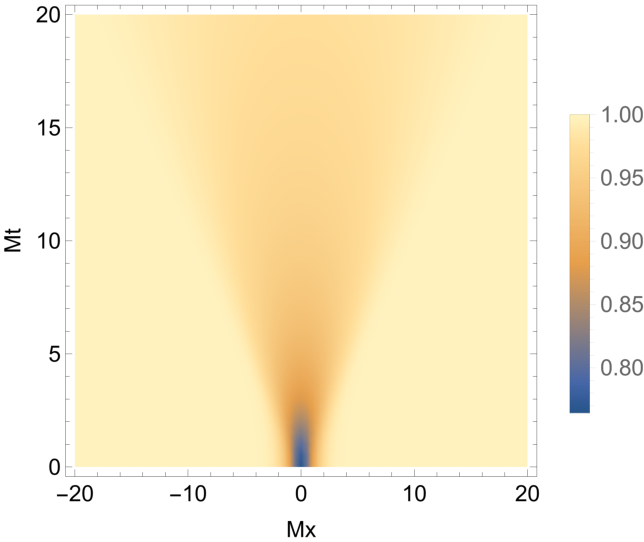}
    \end{subfigure}\hspace{1cm}%
    \begin{subfigure}[h]{0.45\textwidth}
        \includegraphics[width=\textwidth]{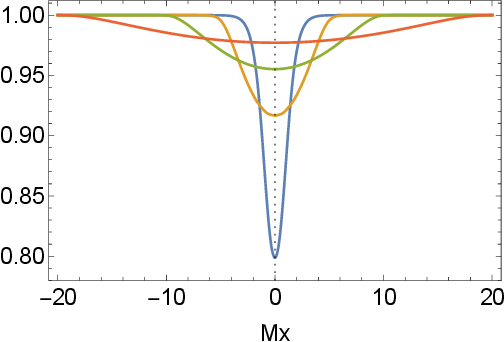}
    \end{subfigure}
    \caption{Potts order parameter component $\langle\sigma_3(x,t)\rangle_{12}/\langle\sigma_3\rangle_{1}$ in the nonequilibrium state (\ref{gauss}) with $MR=1$. On the right, $Mt=1,\,5,\,10,\,20$ in order of increasing value at $x=0$. The deviation from 1 measures the presence of the phase not selected by the initial condition.}
    \label{potts_2}
\end{figure}

Figures~\ref{potts_1} and \ref{potts_2} show the order parameter components $\langle\sigma_1(x,t)\rangle_{12}$ and $\langle\sigma_3(x,t)\rangle_{12}$ in the state (\ref{gauss}). In (\ref{third}) and fig.~\ref{potts_2} the deviation from 1 measures the presence of phase 3. Although this phase is not selected by the initial condition, it is produced by quantum fluctuations.

\subsection{Ashkin-Teller and the unbinding transition}
\label{Ashkin}
The Ashkin-Teller chain \cite{Solyom,BOBD} corresponds to two transverse field Ising chains with site variables $\sigma_{1,i}$ and $\sigma_{2,i}$ interacting via the Hamiltonian
\EQ
H_\textrm{AT}=-J\sum_i\left[\sigma^x_{1,i}\sigma^x_{1,i+1}+\sigma^x_{2,i}\sigma^x_{2,i+1}+\lambda\,\sigma^x_{1,i}\sigma^x_{1,i+1}\sigma^x_{2,i}\sigma^x_{2,i+1}+\,g(\sigma^z_{1,i}+\sigma^z_{2,i}+\lambda\,\sigma^z_{1,i}\sigma^z_{2,i})\right].
\label{AT}
\EN
The theory possesses a $\mathbb{Z}_2$ symmetry in each of the two Ising variables, as well as the symmetry under exchange of the two variables. It is characterized by the fact that $g=1$ leads to a {\it line} of critical points as $\lambda$ varies in an interval including the decoupling point $\lambda=0$, with critical exponents varying continuously with $\lambda$ \cite{Baxter}. In the ferromagnetically ordered regime there are four degenerate ground states $|0_{++}\rangle$, $|0_{+-}\rangle$, $|0_{-+}\rangle$, $|0_{--}\rangle$, labeled by the signs that the two Ising order parameters take in each of them. The vicinity of the critical line is described by the sine-Gordon theory \cite{ZZ}, through a mapping that determines, in particular, the nature of the kinks interpolating between the different ground states \cite{DG}. A pair of ground states such as $|0_{++}\rangle$ and $|0_{-+}\rangle$ ($|0_{+-}\rangle$), related by spin reversal in the first (second) Ising copy, is connected by a kink $A_1$ ($A_2$) with mass $M$. For $\lambda>0$, $A_1$ and $A_2$ form a bound state $B$ with mass $M_B$ that connects the ground states $|0_{++}\rangle$ and $|0_{--}\rangle$. Consider now $\langle\Phi(x,t)\rangle_{++,--}$. We deduced from the analysis of section \ref{sec_main} and the appendix that the leading behavior at large times for $|x|\ll t$ is determined by the $n$-kink states with minimal $n$ among those connecting $|0_{++}\rangle$ and $|0_{--}\rangle$. For $\lambda>0$, $n=1$ and the relevant states are given by (\ref{psi1}) with $|\theta\rangle=|B(\theta)\rangle$. However, when the decoupling point $\lambda=0$ is approached, $M_B$ tends to the unbinding threshold $2M$, and for $\lambda<0$ the bound state $B$ no longer exists. Hence, for $\lambda<0$ the space ${\cal W}_1$ is empty and the dominant contribution comes from the $n=2$ state $A_1A_2$. It follows that two-kink contributions that in the appendix are neglected as subleading with respect to one-kink contributions become leading for $\lambda<0$ and modify the result (\ref{results}) for $|x|\ll t$. This same unbinding mechanism accounts for interfacial wetting  in the theory of phase separation in classical systems at equilibrium \cite{DV,DS,localization,DSS}. Clearly, the mechanism requires at least three degenerate ground states. For the Potts chain of the previous section the existence of a single-kink excitation connecting any pair of ground states is ensured by the permutational symmetry.

\section{Conclusion}
We studied the role of initial conditions in nonequilibrium quantum dynamics in the framework of one-dimensional ferromagnets in the regime of spontaneously broken symmetry. We considered domain wall initial conditions, generally intended as initial conditions that spatially interpolate between two different ground states. The interpolation is arbitrary, with the only constraint of preserving the symmetry characteristic of the equilibrium universality class (e.g. the $\mathbb{Z}_2$ symmetry for Ising). In this setting we obtained analytical results for the one-point functions of local operators at large times. We showed that in this limit the time evolution takes place inside a light cone produced by the spatial inhomogeneity of the initial condition, and that in the innermost region of the light cone ($|x|\ll t$) the space-time dependence is (up to an overall amplitude depending on the initial condition) universal, namely is determined by data of the equilibrium universality class. The origin of the universality is that the result in this region is determined by the excitations with the largest wavelength, which are maximally insensitive to the fine structure of the initial condition. This result should then hold also when the distance from the critical point is not small, in spite of the fact that it was derived in the continuum limit. We also showed that the large time limit curve in the variable $x/t$ (which is nontrivial for operators that distinguish between the two ground states involved in the initial condition) changes with the initial condition. Our formalism also allowed us to show that in systems with more than two degenerate ground states the tuning of an interaction parameter (within the spontaneously broken regime) can change the structure of the space of nonequilibrium states, since the subspace of one-kink excitations disappears via the unbinding of a bound state. The corresponding transition is the nonequilibrium quantum analog of the interfacial wetting transition observed at equilibrium in classical systems at phase coexistence.

\appendix
\section{Composite excitations}
Let us consider the state 
\EQ
|\psi\rangle=|\psi_1\rangle+|\psi_2\rangle\,,
\label{psi1+2}
\EN
where $|\psi_1\rangle$ is the one-kink state (\ref{psi1}) and $|\psi_2\rangle$ is a superposition of two-kink states $|K_{ac}(\theta_1)K_{cb}(\theta_2)\rangle$ with $a,b,c$ all different. Since such a state requires at least three degenerate ground states, we will consider the three-state Potts universality class of section~\ref{sec_potts}. It is convenient to exploit the fact that this model possesses a duality between the ferromagnetically ordered and the paramagnetic regime. One implication is the correspondence $A=K_{a,a+1\,(\textrm{mod}\,3)}$, $\bar{A}=K_{a,a-1\,(\textrm{mod}\,3)}$ between the kinks of the ordered phases and the elementary excitations $A$, $\bar{A}$ of the paramagnetic phase. $A$ and $\bar{A}$ are charge conjugated quasiparticles and the theory is invariant under charge conjugation. It follows that a general superposition of states $|K_{12}K_{23}\rangle$ corresponds to
\EQ
|\psi_2\rangle=\int d\theta_1d\theta_2\,f_2(\theta_1,\theta_2)\,|A(\theta_1)A(\theta_2)\rangle\,.
\label{psi2}
\EN
The Potts theory is integrable in the scaling limit we consider \cite{KS,Zamo_potts}, and we have\footnote{See \cite{CDGJM} for an early advanced application of this formalism in the Potts paramagnetic phase.} 
\bea
&& \hspace{-.5cm}\langle A(\theta_1')\ldots A(\theta_m')|\tilde{\Phi}(0,0)|A(\theta_1)\ldots A(\theta_n)\rangle=\nonumber\\
&& \langle A(\theta_2')\ldots A(\theta_m')|\tilde{\Phi}(0,0)|\bar{A}(\theta_1'+i\pi)A(\theta_1)\ldots A(\theta_n)\rangle+ \sum_{j=1}^n 2\pi\delta(\theta_1'-\theta_j)\left[\prod_{k=1}^{j-1}S_{AA}(\theta_k-\theta_1')\right]\nonumber\\
&& \times\langle A(\theta_2')\ldots A(\theta_m')|\tilde{\Phi}(0,0)|A(\theta_1)\ldots A(\theta_{j-1})A(\theta_{j+1})\ldots A(\theta_n)\rangle\,,
\label{crossing}
\eea
where $S_{AA}(\theta_1-\theta_2)$ is the scattering amplitude\footnote{The scattering in the three-state Potts theory is completely elastic, meaning that the final state is identical to the initial one.} of $A(\theta_1)$ with $A(\theta_2)$; it satisfies crossing
\EQ
S_{AA}(\theta)=S_{\bar{A}A}(i\pi-\theta)\,,
\EN 
and unitarity
\EQ
S_{AA}(\theta)S_{AA}(-\theta)=1\,.
\label{unitarity}
\EN
We also took into account that when working in the paramagnetic phase we have to consider the dual $\tilde{\Phi}$ of the operator $\Phi$ of interest in the regime of spontaneously broken symmetry. Iterative use of (\ref{crossing}) allows one to express any matrix element in terms of the form factors
\EQ
F_{\alpha_1\ldots\alpha_n}^{\tilde{\Phi}}(\theta_1,\ldots,\theta_n)=\langle 0|\tilde{\Phi}(0,0)|\alpha_1(\theta_1)\ldots\alpha_n(\theta_n)\rangle\,,
\label{formfactors}
\EN
where $\alpha_i=A,\bar{A}$, and $|0\rangle$ is the unique ground state of the paramagnetic phase. We will consider operators whose expectation values $\langle\Phi\rangle_a$ in the ordered phases are $a$-independent. This introduces some simplifications in the equations satisfied by the form factors, which read \cite{Smirnov}
\EQ
F_{\ldots\alpha_j\alpha_{j+1}\ldots}^{\tilde{\Phi}}(\ldots,\theta_j,\theta_{j+1},\ldots)=S_{\alpha_j\alpha_{j+1}}(\theta_j-\theta_{j+1})F_{\ldots\alpha_{j+1}\alpha_j\ldots}^{\tilde{\Phi}}(\ldots,\theta_{j+1},\theta_j,\ldots)\,,
\label{ff1}
\EN
\EQ
F_{\alpha_1\ldots\alpha_n}^{\tilde{\Phi}}(\theta_1+2i\pi,\theta_2,\ldots,\theta_n)=F_{\alpha_2\ldots\alpha_n,\alpha_1}^{\tilde{\Phi}}(\theta_2,\ldots,\theta_n,\theta_1)\,,
\label{ff2}
\EN
\bea
&& \hspace{-1cm}\textrm{Res}_{\theta'=\theta}F_{\bar{\alpha}\beta\alpha_1\ldots\alpha_n}^{\tilde{\Phi}}(\theta'+i\pi,\theta,\theta_1,\ldots,\theta_n)\nonumber\\
&& =i\delta_{\alpha\beta}\left[1-\prod_{j=1}^{n}S_{\alpha\alpha_j}(\theta-\theta_j)\right]
F_{\alpha_1\ldots\alpha_n}^{\tilde{\Phi}}(\theta_1,\theta_2,\ldots,\theta_n)\,.
\label{ff3}
\eea

Let us consider $\langle\Phi(x,t)\rangle_{13}=\langle\psi|\Phi(x,t)|\psi\rangle/\langle\psi|\psi\rangle$ with $|\psi\rangle$ given by (\ref{psi1+2}). The contribution proportional to $\langle\psi_1|\Phi|\psi_1\rangle$ follows from the results of section~\ref{sec_main}. We now consider the contribution proportional to
\bea
\langle\psi_2|\Phi(x,t)|\psi_2\rangle &=&\int d\theta_1 d\theta_2 d\theta_3 d\theta_4\,f_2^*(\theta_2,\theta_1)f_2(\theta_3,\theta_4)\label{phi22}\\
&\times & \langle A(\theta_2)A(\theta_1)|\tilde{\Phi}(0,0)|A(\theta_3)A(\theta_4)\rangle\,
e^{i[(p_1+p_2-p_3-p_4)x+(E_1+E_2-E_3-E_4)t]}\,,\nonumber
\eea
where
\bea
&& \hspace{-2cm}\langle A(\theta_2)A(\theta_1)|\tilde{\Phi}(0,0)|A(\theta_3)A(\theta_4)\rangle\nonumber\\
&=& F^{\tilde{\Phi}}_{\bar{A}AA\bar{A}}(\theta_2+i\pi,\theta_3,\theta_4,\theta_1-i\pi)\nonumber\\
&+& 2\pi\left[\delta(\theta_{14})S_{AA}(\theta_{12})S_{AA}(\theta_{31})F^{\tilde{\Phi}}_{\bar{A}A}(\theta_2+i\pi,\theta_3)+\delta(\theta_{13})S_{AA}(\theta_{12})F^{\tilde{\Phi}}_{\bar{A}A}(\theta_2+i\pi,\theta_4) \right. \nonumber\\
&& \hspace{.2cm}+\left.\delta(\theta_{23})F^{\tilde{\Phi}}_{\bar{A}A}(\theta_1+i\pi,\theta_4)+\delta(\theta_{24})S_{AA}(\theta_{34})F^{\tilde{\Phi}}_{\bar{A}A}(\theta_1+i\pi,\theta_3) \right]\nonumber\\
&+& (2\pi)^2\left[\delta(\theta_{23})\delta(\theta_{14})+\delta(\theta_{24})\delta(\theta_{13})S_{AA}(\theta_{32})\right]\langle\tilde{\Phi}\rangle\,,
\label{M_phi}
\eea
with $\theta_{ij}=\theta_i-\theta_j$ and $\langle\tilde{\Phi}\rangle=\langle 0|\tilde{\Phi}|0\rangle=\langle\Phi\rangle_a$. 

Let us call $G_4$ the contribution to (\ref{phi22}) of the term $F^{\tilde{\Phi}}_{\bar{A}AA\bar{A}}$ in (\ref{M_phi}). It follows from (\ref{ff3}) that when integrating over $\theta_2$ we have to deal with poles at $\theta_2=\theta_3,\theta_4$. Proceeding as in section~\ref{sec_main}, the contribution $G_4^{\textrm{pole}}$ of these poles at large times is determined by the residues on the poles, which we know from (\ref{ff3}), and reads
\bea
&& G_4^{\textrm{pole}}(x,t)\simeq -2\pi\int_{\theta_0}^\infty d\theta\label{g4pole}\\
&& \hspace{-1cm}\left\{\int d\theta_1 d\theta_4 \,f_2^*(\theta,\theta_1)f_2(\theta,\theta_4)F^{\tilde{\Phi}}_{A\bar{A}}(\theta_4,\theta_1-i\pi)\left[1-S_{AA}(\theta_1-\theta)S_{AA}(\theta-\theta_4)\right]e^{i[p_{14}x+E_{14}t]}+\right.\nonumber\\
&& \hspace{-1cm}\left.\int d\theta_1 d\theta_3 \,f_2^*(\theta,\theta_1)f_2(\theta_3,\theta)S_{AA}(\theta_3-\theta)F^{\tilde{\Phi}}_{A\bar{A}}(\theta_3,\theta_1-i\pi)\left[1-S_{AA}(\theta_1-\theta)S_{AA}(\theta-\theta_3)\right]e^{i[p_{13}x+E_{13}t]}\right\}\,,\nonumber
\eea
where $p_{ij}=p_i-p_j$ and $E_{ij}=E_i-E_j$. Since (\ref{ff3}) shows that $F^{\tilde{\Phi}}_{A\bar{A}}(\theta_4,\theta_1-i\pi)$ has no pole\footnote{This corresponds to the fact that (\ref{expansion}) has no pole for $\langle\Phi\rangle_a=\langle\Phi\rangle_b$.} on the integration path, the behavior of (\ref{g4pole}) at large times can now be analyzed as the contribution of regular terms along the lines already seen in section~\ref{sec_main}. The stationary phase condition yields the light cone and the suppression of the integral outside it. Deeply inside the light cone, namely for $|x|/t\ll 1$, small values of $\theta_1,\theta_4$ ($\theta_1,\theta_3$) dominate in the first (second) term, and the expressions in the square brackets become $1-S_{AA}(-\theta)S_{AA}(\theta)$, which vanishes due to (\ref{unitarity}). Hence, $G_4^{\textrm{pole}}$ can be ignored in the regions specified in (\ref{results}). Concerning the contribution $G_4^{\textrm{reg}}$ coming from the regular part of $F^{\tilde{\Phi}}_{\bar{A}AA\bar{A}}$, we have again suppression outside the light cone and dominance of small rapidities $\theta_1,\ldots,\theta_4$ for $|x|/t\ll 1$. In this region, rescaling of rapidities in (\ref{phi22}) yields that  $G_4^{\textrm{reg}}$ is suppressed at least as $t^{-2}$, and is then subleading with respect to the one-kink contribution.

Let us now call $G_2$ the contribution to (\ref{phi22}) of the four terms in (\ref{M_phi}) containing $F^{\tilde{\Phi}}_{\bar{A}A}$. It will be sufficient to consider one of these terms, say $\delta(\theta_{14})S_{AA}(\theta_{12})S_{AA}(\theta_{31})F^{\tilde{\Phi}}_{\bar{A}A}(\theta_2+i\pi,\theta_3)$. Since the form factor has no pole at $\theta_2=\theta_3$, we have suppression of the integral outside the light cone and dominance of small values of $\theta_2,\theta_3$ for $|x|/t\ll 1$. In this region $f_2^*(\theta_2,\theta_1)f_2(\theta_3,\theta_1)S_{AA}(\theta_{12})S_{AA}(\theta_{31})F^{\tilde{\Phi}}_{\bar{A}A}(\theta_2+i\pi,\theta_3)\simeq |f_2(0,\theta_1)|^2S_{AA}(\theta_{1})S_{AA}(-\theta_{1})F^{\tilde{\Phi}}_{\bar{A}A}(i\pi,0)$, which reduces to $|f_2(0,\theta_1)|^2C_0^\Phi$ using (\ref{expansion}), (\ref{unitarity}) and duality. The integral over $\theta_2$ and $\theta_3$ is analogous to (\ref{Greg}) and, taking into account that the other terms in $G_2$ behave in the same way, we get
\EQ
G_2(x,t)\simeq B_{f_2}\frac{C_0^\Phi}{Mt}\,,\hspace{1cm}|x|/t\ll 1\,.
\EN

The last contribution to (\ref{phi22}) comes from the term in (\ref{M_phi}) proportional to $\langle\tilde{\Phi}\rangle$, and is equal to $G_0=\langle\psi_2|\psi_2\rangle\,\langle\tilde{\Phi}\rangle=\langle\psi_2|\psi_2\rangle\,\langle\Phi\rangle_a$. 

Finally, $\langle\Phi(x,t)\rangle_{13}$ includes the off-diagonal contribution proportional to $\langle\psi_1|\Phi|\psi_2\rangle+\langle\psi_2|\Phi|\psi_1\rangle$. It is sufficient to consider the first term, which involves 
\EQ
\langle \bar{A}(\theta_1)|\tilde{\Phi}(0,0)|A(\theta_2)A(\theta_3)\rangle=F^{\tilde{\Phi}}_{AAA}(\theta_1+i\pi,\theta_2,\theta_3)\,.
\EN
Since (\ref{ff3}) shows that this matrix element yields no poles on the integration path, we have suppression outside the light cone and dominance of small $\theta_1,\theta_2,\theta_3$ for $|x|/t\ll 1$. In this region rescaling of the rapidities shows at least a $t^{-3/2}$ suppression at large times, which is again subleading with respect to the one-kink contribution. 

Putting all together, and recalling that $\langle\psi|\psi\rangle=\langle\psi_1|\psi_1\rangle+\langle\psi_2|\psi_2\rangle$, we see that inclusion in $|\psi\rangle$ of the two-kink contribution gives again the result (\ref{results}), specialized to the case of $a$-independent $\langle\Phi\rangle_a$ that we considered in this appendix. The difference with respect to the one-kink result is a change of the constant ${\cal A}$ that encodes the dependence on the initial condition.



\end{document}